\documentclass[12pt]{article}
\usepackage{oldgerm}
\usepackage{yfonts}
\usepackage{euler}
\usepackage{graphics}
\usepackage{bm}
\usepackage{graphicx}
\usepackage{epstopdf}
\usepackage{amsmath}
\usepackage{amssymb}
\usepackage{amstext}
\usepackage{amscd}
\usepackage{amsfonts}
\usepackage{appendix}
\usepackage{color}
\DeclareMathAlphabet{\EuFrak}{U}{euf}{m}{n}
\DeclareMathAlphabet{\EuScript}{U}{eus}{m}{n}

\newcommand{\nd}{\noindent}

\title{{\bf Quantum q-Field Theory: q-Schr\"{o}dinger
and q-Klein-Gordon Fields }}

\author{{A. Plastino$^1$and M. C. Rocca$^1$}\\
\small{$^1$ La Plata National University and   Argentina's National Research Council}\\
\small{(IFLP-CCT-CONICET)-C. C. 727, 1900 La Plata - Argentina}}

\date{\today}

\begin{document}

\maketitle

\begin{abstract}

We show how to deal with  the generalized q-Schr\"{o}dinger and 
q-Klein-Gordon fields in a  variety of scenarios. These 
q-fields are meaningful at very high energies (TeVs) for 
for $q=1.15$, high ones  (GeVs)
for $q=1.001$, and low energies (MeVs)for  $q=1.000001$ [Nucl. Phys. A {\bf 948}  (2016) 19; 
Nucl. Phys. A {\bf 955}  (2016) 16].
(See the Alice experiment of LHC).
We develop here the quantum field theory (QFT) for the q-Schr\"{o}dinger and 
 q-Klein-Gordon fields, 
 showing that both reduce to the customary Schr\"{o}dinger and 
 Klein-Gordon QFTs  for q close to unity. 
 Further, we analyze the q-Klein-Gordon field for $q\geq 1.15$ . In this case for
$2q-1=n$ (n integer $\geq 2$) and analytically compute the self-energy and the propagator up to second order.

\nd {\bf Keywords:}Non-linear Klein-Gordon equation;
Non-linear Schr\"{o}dinger equation;
Classical Field Theory, Quantum Field Theory.\\
\nd {\bf PACS:} 11.10.Ef; 11.10.Lm; 02.30.Jm

\end{abstract}

\newpage

\renewcommand{\theequation}{\arabic{section}.\arabic{equation}}

\setcounter{equation}{0}

\section{Introduction}

Classical fields theories (CFT) associated to  Tsallis' q-scenarios have 
been intensely studied recently
 \cite{tp1,44,ts5}. Associated quantum (QFT) treatments
have also been discussed \cite{ts5}. In this paper we show how to treat the q-Schr\"{o}dinger and 
q-Klein-Gordon fields in a variety of cases. It has been shown in \cite{ts1,ts2} that 
q-fields emerge at1)  very high energies
(TeV) for $q=1.15$, 2) High (GeV)
for $q=1.001$, and 3) low (MeV) for $q=1.000001$.
LHC-Alice experiments show that Tsallis q-effects manifest themselves  \cite{ts3}
at TeV energies. 

In this effort we develop QFTs associated to q-Schr\"{o}dinger and q-Klein-Gordon
fields. Moreover, we study the q-KG field in the case 
$2q-1=n,\,\, n$ integer $\geq 2$. Here we evaluate the selfenergy and propagator up to ssecond order, thus generalizing results of  \cite{ts5}.
In this respect, note also recent work on Proca-de Broglies'
classical field theory \cite{narp}.

Motivations for nonlinear quantum evolution equations can be divided up into two types,   namely,  (A) as basic equations governing  phenomena
at the frontiers of quantum mechanics, mainly 
at the boundary between quantum
 and gravitational physics (see, [\cite{37,38}
and references therein). The other possibility is (B) regard nonlinear-Schr\"{o}dinger-like equations (NLSE)
 as effective, single particle mean field
descriptions of involved  quantum many-body systems. A paradigmatic illustration  is that of \cite{39}. 
In earlier applications
of nonlinear Schr\"{o}dinger equations, one encounters situations
 involving a cubic nonlinearity in
the wave function.

Referring to (A), our  present  NLSE can be used for
a description of dark matter components, since the associated 
 variational principle (the one that  leads to the NLSE) 
 is seen to describe particles that can not interact
with the electromagnetic field \cite{tp0}.  Withe reference to  (B), we remark that
 the NLSE displays strong similarity with
the Schr\"{o}dinger equation linked to a particle endowed with
a time-position dependent effective mass \cite{40,41,42,43} , involving
  particles moving in nonlocal potentials, reminiscent of the energy density functional quantum
many-body problem's approach  \cite{44}.

During the last years, the search for insight into a number of complex phenomena produced 
 interesting proposals involving localized solutions attached to non linear Klein-Gordon and Schr\"{o}dinger equations, i.e., non linear generalizations of these equations \cite{tp0,tp1}.  Following \cite{tp0}, we extend these generalizations  here by developing quantum field theories (QFT)  
 associated to the q-Schr\"{o}dinger and 
 q-Klein-Gordon equations \cite{tp1}.

Here, we develop first the classical field theory (CFT) associated to that
 q-Schr\"{o}dinger equation
deduced in \cite{tp2} from the hypergeometric differential equation.
We define the corresponding physical fields via an analogy with treatments in string theory \cite{tp4}
for defining physical states of the bosonic string.
Our ensuing theory reduces to the conventional Schr\"{o}dinger  field theory for
$q\rightarrow 1$.

Secondly, we develop the QFT for that very q-Schr\"{o}dinger equation 
(see also  \cite{tp3}). {\it This} equation  is similar but not identical to that advanced in
 \cite{tp1}. Its treatment is however much simpler than that employed in \cite{tp0}.

In the third place, we develop the QFT for the q-K-G Field
in several scenarios, generalizing results  of 
 \cite{ts5} and showing that the ensuing q-K-G field  reduces to 
the customary  K-G field for $q\rightarrow 1$.

\setcounter{equation}{0}

\section{A non-linear q-Schr\"{o}dinger  Equation}

\subsection{Classical Theory}

We develop here the CFT for that  particular q-Schr\"{o}dinger  Equation
advanced in \cite{tp3} from the Hypergeometric Differential Equation. This NLSE is different from the pioneer one
proposed in \cite{tp1}, but exhibits better qualitative features. One has
 
 \begin{equation}
\label{eq2.1}
i\hbar \frac {\partial}{\partial t}\psi(\vec{x},t)^q=
H\psi(\vec{x},t).
\end{equation}
In  the free particle instance one writes
\begin{equation}
\label{eq2.2}
H_0=-\frac {\hbar^2} {2m}\bigtriangleup,
\end{equation}
whose solution reads

\begin{equation}
\label{eq2.3}
\psi(\vec{x},t)=[1+(1-q)\frac {i} {\hbar}
(\vec{p}\cdot\vec{x}-Et)]^{\frac {1} {1-q}}.
\end{equation}
Introduce now  action

\[{\cal S}=\frac {1} {(4q-2)V}\int\limits_{-\infty}^{\infty}
\int\limits_V\left(i\hbar \psi^{\dagger q}\partial_t\phi^{\dagger}-
i\hbar \psi^{q}\partial_t\phi-
\frac {\hbar^2} {2m}\nabla\psi\nabla\phi-\right.\]
\begin{equation}
\label{eq2.4}
\left.\frac {\hbar^2} {2m}\nabla\psi^{\dagger}\nabla\phi^{\dagger}\right)
dt\;d^3x,
\end{equation}
with  $V$ the Euclidian volume. Our action can be rewritten in the fashion

\begin{equation}
\label{eq2.5}
{\cal S}=\int\limits_{-\infty}^{\infty}
\int\limits_V
{\cal L}(\psi,\psi^{\dagger},\partial_t\phi,
\partial_t\phi^{\dagger},\nabla\psi,\nabla\psi^{\dagger},
\nabla\phi,\nabla\phi^{\dagger})dt\;d^3x.
\end{equation}
One obtains from  (\ref{eq2.5}) the field's motion equations

\begin{equation}
\label{eq2.6}
i\hbar \frac {\partial}{\partial t}\psi^q(\vec{k},t)+
\frac {\hbar^2} {2m}\bigtriangleup\psi(\vec{x},t)=0,
\end{equation}
\begin{equation}
\label{eq2.7}
i\hbar q\psi(\vec{x},t)^{q-1} \frac {\partial}{\partial t}\phi(\vec{k},t)-
\frac {\hbar^2} {2m}\bigtriangleup\phi(\vec{x},t)=0.
\end{equation}
whose  solution is (\ref{eq2.3}). Instead, that for
 (\ref{eq2.7}) reads 

\begin{equation}
\label{eq2.8}
\phi(\vec{x},t)=[1+(1-q)\frac {i} {\hbar}
(\vec{p}\cdot\vec{x}-Et)]^{\frac {2q-1} {q-1}}.
\end{equation}
If  $q\rightarrow 1$, $\phi$ becomes  $\psi^{\dagger}$,
the adjoint of  $\psi$. Now, the concomitant canonically conjugated momenta are

\[\Pi_{\psi}=\frac {\partial{\cal L}} {\partial(\partial_t\psi)}=0\;\;\;;\;\;\;
\Pi_{\psi^{\dagger}}=\frac {\partial{\cal L}} {\partial(\partial_t\psi^{\dagger})}=0\],
\begin{equation}
\label{eq2.9}
\Pi_{\phi}=\frac {\partial{\cal L}} {\partial(\partial_t\phi)}=
-\frac {i\hbar\psi^q} {(4q-2)V}\;\;\;;\;\;\;
\Pi_{\phi^{\dagger}}=\frac {\partial{\cal L}} {\partial(\partial_t\phi^{\dagger})}=
\frac {i\hbar\psi^{\dagger q}} {(4q-2)V},
\end{equation}
and the associated Hamiltonian is

\begin{equation}
\label{eq2.10}
{\cal H}=\Pi_{\phi}\partial_t\phi+\Pi_{\phi^{\dagger}}\partial_t\phi^{\dagger}-
{\cal L},
\end{equation}
that we cast in terms of  $\psi$ - $\phi$    as

\begin{equation}
\label{eq2.11}
{\cal H}=\frac {\hbar^2} {(8q-4)mV}
\left(\nabla\psi\nabla\phi+\nabla\psi^{\dagger}\nabla\phi^{\dagger}\right).
\end{equation}
The field energy is
\begin{equation}
\label{eq2.12}
E=\int\limits_V{\cal H}d^3x.
\end{equation}
If we replace the solutions  (\ref{eq2.3}) and (\ref{eq2.8}) into
(\ref{eq2.12}), one has

\begin{equation}
\label{eq2.13}
E=\int\limits_V\frac {\hbar^2} {(8q-4)mV}(4q-2)\frac {p^2}
{\hbar^2}d^3x,
\end{equation}
or
\begin{equation}
\label{eq2.14}
E=\frac {p^2} {2m},
\end{equation}
that exactly correspond to the wave energy (\ref{eq2.3}),
as one should expected. The field-momentum density reads

\begin{equation}
\label{eq2.15}
\vec{{\cal P}}=
-\frac {\partial{\cal L}} {\partial(\partial_t\psi)}\nabla\psi
-\frac {\partial{\cal L}} {\partial(\partial_t\phi)}\nabla\phi
-\frac {\partial{\cal L}} {\partial(\partial_t\psi^{\dagger})}\nabla\psi^{\dagger}
-\frac {\partial{\cal L}} {\partial(\partial_t\phi^{\dagger})}\nabla\phi^{\dagger},
\end{equation}
or
\begin{equation}
\label{eq2.16}
\vec{{\cal P}}=
\frac {i\hbar} {(4q-2)V}\left(\psi^q\nabla\phi-
\psi^{\dagger q}\nabla\phi^{\dagger}\right),
\end{equation}
the field-momentum becoming
\begin{equation}
\label{eq2.17}
\vec{P}=\int\limits_V\vec{{\cal P}}d^3x.
\end{equation}
Employing  (\ref{eq2.3}) and (\ref{eq2.8}),
one finds for  the momentum
\begin{equation}
\label{eq2.18}
\vec{P}=\frac {i\hbar} {(4q-2)V}\int\limits_V
\frac {4q-2} {i\hbar}\vec{p}d^3x,
\end{equation}
or
\begin{equation}
\label{eq2.19}
\vec{P}=\vec{p}.
\end{equation}
The probability density is now
\begin{equation}
\label{eq2.20}
\rho=\frac {1} {2V} [\psi^q\phi+\psi^{\dagger q}\phi^{\dagger}],
\end{equation}
verifying
\begin{equation}
\label{eq2.21}
\frac {\partial} {\partial t}\rho+\nabla\cdot\vec{j}=K,
\end{equation}
where
\begin{equation}
\label{eq2.22}
\vec{j}=\frac {\hbar^2(q+1)} {8mVqi}[\phi\nabla\psi-
\psi\nabla\phi+\psi^{\dagger}\nabla\phi^{\dagger}-
\phi^{\dagger}\nabla\psi^{\dagger}],
\end{equation}
is the probability-current.  $K$ reads
\begin{equation}
\label{eq2.23}
K=\frac {\hbar^2(q-1)} {8mVqi}[
\psi^{\dagger}\bigtriangleup\phi^{\dagger}+
\phi^{\dagger}\bigtriangleup\psi^{\dagger}-
\phi\bigtriangleup\psi-\psi\bigtriangleup\phi],
\end{equation}
that vanishes at  $q=1$.
However, the {\it physical} fields are those for which $K=0$. For example, one lists as physical 
the solutions  (\ref{eq2.3}) and (\ref{eq2.8}), since for them probability 
is indeed conserved.

\subsection{Quantum Theory}

We start with the action
\[{\cal S}=-\int
\left(i\hbar \psi^{q}\partial_t\phi-
i\hbar \psi^{\dagger q}\partial_t\phi^{\dagger} +
\frac {\hbar^2} {2m}\nabla\psi\nabla\phi+\right.\]
\begin{equation}
\label{eq2.24}
\left.\frac {\hbar^2} {2m}\nabla\psi^{\dagger}\nabla\phi^{\dagger}\right)
dt\;d^3x. 
\end{equation}
We develop first a theory for 1)  $q$ close to unity and 2) weak fields $\psi$. In these condiitions one appeals to the approximation
\begin{equation}
\label{eq2.25}
\psi^q\simeq\psi+(q-1)\psi\ln\psi,
\end{equation}
and since $\psi$ is a weak field
\begin{equation}
\label{eq2.26}
\psi\simeq I+(q-1)\eta.
\end{equation}
Consequently, the action  (\ref{eq2.24})  becomes
\[{\cal S}=-(q-1)\int
\left(i\hbar \eta\partial_t\phi -
i\hbar \eta^{\dagger}\partial_t\phi^{\dagger}+
\frac {\hbar^2} {2m}\nabla\eta\nabla\phi+\right.\]
\begin{equation}
\label{eq2.27}
\left.\frac {\hbar^2} {2m}\nabla\eta^{\dagger}\nabla\phi^{\dagger}\right)
dt\;d^3x,
\end{equation}
where we used
\begin{equation}
\label{eq0.1}
\int\eta(\vec{x},t)dtd^3x=
\int\phi(\vec{x},t)dtd^3x=0,
\end{equation}
since the fields are 
\begin{equation}
\label{eq2.28}
\eta(\vec{x},t)=\frac {1} {(2\pi\hbar)^{\frac {3} {2}}}
\int a(\vec{p})
e^{\frac {i} {\hbar}(\vec{p}\cdot\vec{x}-Et)}d^3p,
\end{equation}
(see \cite{ts4}) 
\begin{equation}
\label{eq2.29}
\eta^{\dagger}(\vec{x},t)=\frac {1} {(2\pi\hbar)^{\frac {3} {2}}}
\int a^{\dagger}(\vec{p})
e^{-\frac {i} {\hbar}(\vec{p}\cdot\vec{x}-Et)}d^3p
\end{equation}
\begin{equation}
\label{eq2.30}
\phi(\vec{x},t)=\frac {1} {(2\pi\hbar)^{\frac {3} {2}}}
\int b(\vec{p})
e^{\frac {i} {\hbar}(\vec{p}\cdot\vec{x}-Et)}d^3p,
\end{equation} and 
\begin{equation}
\label{eq2.31}
\phi^{\dagger}(\vec{x},t)=\frac {1} {(2\pi\hbar)^{\frac {3} {2}}}
\int b^{\dagger}(\vec{p})
e^{-\frac {i} {\hbar}(\vec{p}\cdot\vec{x}-Et)}d^3p.
\end{equation}
{\it Surprisingly enough, the q-Schr\"{o}dinger field (qSF)
reduce to the usual SF of low energies! }
Creation-destruction operators verify
\begin{equation}
\label{eq0.2}
[a(\vec{p}),a^{\dagger}(\vec{p}^{'})]=[b(\vec{p}),b^{\dagger}(\vec{p}^{'})]=
\delta(\vec{p}-\vec{p}^{'}).
\end{equation}
The propagator for the field  $\eta$ is \cite{ts4}
\begin{equation}
\label{eq2.32}
\Delta_{\eta}(\vec{x},t)=\left(\frac {m} {2\pi i \hbar}\right)^{
\frac {3} {2}}
t_+^{-\frac {3} {2}}
e^{\frac {im\vec{x}^2} {2\hbar t}},
\end{equation}
that, in terms of energy and momentum reads
\begin{equation}
\label{eq2.33}
\hat{\Delta}_{\eta}(\vec{p},E)=
\frac {i\hbar} {E-\frac {\vec{p}^2} {2m}+i0}.
\end{equation}
These two representations  are related via 
\begin{equation}
\label{eq2.34}
\Delta_{\eta}(\vec{x},t)=\frac {1} {(2\pi\hbar)^4}
\int\hat{\Delta}(\vec{p},E)
e^{\frac {i} {\hbar}(\vec{p}\cdot\vec{x}-Et)}dEd^3p.
\end{equation}
The convolution of this propagator with itself,
 with $E$ and $\vec{p}$ as variables,  is NOT finite. It can be calculated, however, by appeal to distributions' theory  using the relation
\begin{equation}
\label{eq2.35}
\hat{f}\ast\hat{g}=(2\pi\hbar)^4{\cal F}(fg).
\end{equation}
This is so because divergences in the convolution of two 
phase space functions derive from multiplication of distributions possessing singularities 
{\it at the same} configuration-space point. Keeping in mind that 
\begin{equation}
\label{eq2.36}
\Delta_{\eta}^2(\vec{x},t)
=\left(\frac {m} {2\pi i \hbar}\right)^3
t_+^{-3}e^{\frac {im\vec{x}^2} {\hbar t}},
\end{equation}
(\ref{eq2.35}) yields
\begin{equation}
\label{eq2.37}
\frac{1} {(2\pi\hbar)^4}\left(
\hat{\Delta}_{\eta}(\vec{p},E)\ast\hat{\Delta}_{\eta}(\vec{p},E)\right)=
\int
\left(\frac {m} {2\pi i \hbar}\right)^3t_+^{-3}
e^{\frac {im\vec{x}^2} {\hbar t}}
e^{-\frac {i} {\hbar}(\vec{p}\cdot\vec{x}-Et)}dtd^3x.
\end{equation}
The spatial integral is 
\begin{equation}
\label{eq2.38}
\int
e^{\frac {im\vec{x}^2} {\hbar t}}
e^{-\frac {i} {\hbar}(\vec{p}\cdot\vec{x})}d^3x=
\pi^{\frac {3} {2}}\frac {(i\hbar t)^{\frac {3} {2}}}
{m^{\frac {3} {2}}}
e^{-\frac {i\vec{p}^2t} {4\hbar m}},
\end{equation}
so that the convolution becomes 
\begin{equation}
\label{eq2.39}
\hat{\Delta}_{\eta}(\vec{p},E)\ast\hat{\Delta}_{\eta}(\vec{p},E)=
\frac {(2\pi\hbar)^4} {8}
\left(\frac {m} {i\pi\hbar}\right)^{\frac {3} {2}}
\int
t_+^{-\frac {3} {2}}
e^{\frac {i} {\hbar}(E-\frac {\vec{p}^2} {4m})t}dt.
\end{equation}
Using the result below (see \cite{tp5})
\begin{equation}
\label{eq2.40}
{\cal F}[x_+^{\lambda}]=i
e^{\frac {i\pi\lambda} {2}}\Gamma(\lambda+1)
(k+i0)^{-\lambda-1},
\end{equation}
one finds
\begin{equation}
\label{eq2.41}
\hat{\Delta}_{\eta}(\vec{p},E)\ast\hat{\Delta}_{\eta}(\vec{p},E)=
4\pi^2\hbar^2m^{\frac {3} {2}}
\left(E-\frac {\vec{p}^2} {4m}+i0\right)^{\frac {1} {2}}.
\end{equation}

\setcounter{equation}{0}

\section{A non-linear q-Klein-Gordon Equation}

The classical FT associated to the  
 q-Klein-Gordon equation was developed in 
\cite{ts5}. Here we tackle the quantum version, whose action is  
\[{\cal S}=\int\left\{
\partial_{\mu}\phi(x_{\mu})
\partial^{\mu}\psi(x_{\mu})+
\partial_{\mu}\phi^{\dagger}(x_{\mu})
\partial^{\mu}\psi^{\dagger}(x_{\mu})\right.\]
\begin{equation}
\label{eq3.1}
\left.-qm^2\left[
\phi^{2q-1}(x_{\mu})\psi(x_{\mu})+
\phi^{\dagger 2q-1}(x_{\mu})\psi^{\dagger}(x_{\mu})\right]\right\}d^4x.
\end{equation}
This theory is 1) adequate for very energetic (TeVs)  q-particles, according to CERN-Alice experiments, 
and 2) non re-normalizable  for any $q>1$. Thus, it cannot be dealt with neither with  
dimensional regularization nor with differential one.  A way out is provided by the ultradistributions' convolution of Bollini and Rocca
\cite{tp6,tp7,tp8,tp9}. Ultradistribuions provides a general formalism
to treat non-renormalizable theories and gives in the configuration space
a general product in a ring with zero divisors (a product of distributions
of exponential type). For example we can treat cases with $q\geq 1.15$
as we will do later.

The concomitant theory is tractable here  for  weak fields 
and for A) $q \sim 1$ or B) particular q-values. 
We analyze first the case  $q \sim 1$, associated to energies smaller that 1 TeV. We can thus write
\begin{equation}
\label{eq3.2}
qm^2\phi^{2q-1}=qm^2\phi+2(q-1)m^2\phi\ln\phi.
\end{equation}
Since the field is weak we have
\begin{equation}
\label{eq3.3}
\phi\simeq I+(q-1)\eta,
\end{equation}
\begin{equation}
\label{eq3.4}
\ln\phi\simeq(q-1)\eta,
\end{equation}
Using (\ref{eq3.2}), (\ref{eq3.3}),  and   (\ref{eq3.4}) the field's action becomes
\[{\cal S}=(q-1)\int\left\{
\partial_{\mu}\eta(x_{\mu})
\partial^{\mu}\psi(x_{\mu})+
\partial_{\mu}\eta^{\dagger}(x_{\mu})
\partial^{\mu}\psi^{\dagger}(x_{\mu})\right.\]
\begin{equation}
\label{eq3.5}
\left.-(3q-2)m^2\left[
\eta(x_{\mu})\psi(x_{\mu})+
\eta^{\dagger}(x_{\mu})\psi^{\dagger}(x_{\mu})\right]\right\}d^4x,
\end{equation}
where we employed
\begin{equation}
\label{eq3.6}
\int\eta(x_{\mu})d^4x=
\int\psi(x_{\mu})d^4x=0.  \end{equation}
Defining
\begin{equation}
\label{eq3.7}
3q-2\neq 0\:\:\:\mu^2=(3q-2)m^2.
\end{equation}
one has
\[{\cal S}=(q-1)\int\left\{
\partial_{\mu}\eta(x_{\mu})
\partial^{\mu}\psi(x_{\mu})+
\partial_{\mu}\eta^{\dagger}(x_{\mu})
\partial^{\mu}\psi^{\dagger}(x_{\mu})\right.\]
\begin{equation}
\label{eq3.8}
\left.-\mu^2\left[
\eta(x_{\mu})\psi(x_{\mu})+
\eta^{\dagger}(x_{\mu})\psi^{\dagger}(x_{\mu})\right]\right\}d^4x.
\end{equation}
{\it The low energy field is just the usual Klein-Gordon one!}For the fields we have
\begin{equation}
\label{eq3.9}
\eta(x_{\mu})=\frac {1} {(2\pi)^{\frac {3} {2}}}
\int\left[\frac {a(\vec{k})} {\sqrt{2\omega}}
e^{-ik_{\mu}x^{\mu}}+
\frac {b^{\dagger}(\vec{k})} {\sqrt{2\omega}}
e^{ik_{\mu}x^{\mu}}\right]d^3k,
\end{equation}
\begin{equation}
\label{eq3.10}
\psi(x_{\mu})=\frac {1} {(2\pi)^{\frac {3} {2}}}
\int\left[\frac {c(\vec{k})} {\sqrt{2\omega}}
e^{-ik_{\mu}x^{\mu}}+
\frac {d^{\dagger}(\vec{k})} {\sqrt{2\omega}}
e^{ik_{\mu}x^{\mu}}\right]d^3k,
\end{equation}
where $k_0=\omega=\sqrt{\vec{k}^2+\mu^2}.$\\
Field quantization proceeds then along familiar lines:
\[[a(\vec{k}),a^{\dagger}(\vec{k^{'}})]=[b(\vec{k}),b^{\dagger}(\vec{k^{'}})]=
[c(\vec{k}),c^{\dagger}(\vec{k^{'}})]=\]
\begin{equation}
\label{eq3.11}
[d(\vec{k}),d^{\dagger}(\vec{k^{'}})]=
\delta(\vec{k}-\vec{k^{'}}).
\end{equation}
For  $3q-2=0$,  i.e.,  $q=\frac {2} {3}$,
the low energy theory is one for a null mass field
\begin{equation}
\label{eq3.12}
{\cal S}=-\frac {1} {3}\int\left[
\partial_{\mu}\eta(x_{\mu})
\partial^{\mu}\psi(x_{\mu})+
\partial_{\mu}\eta^{\dagger}(x_{\mu})
\partial^{\mu}\psi^{\dagger}(x_{\mu})\right]d^4x,
\end{equation}
where $k_0=\omega=|\vec{k}|$.

\nd We tackle now the q-KG theory for an integer $n$ such that   
$2q-1=n$, for $m$ small, where the action is 
\[{\cal S}=\int\left\{
\partial_{\mu}\phi(x_{\mu})
\partial^{\mu}\psi(x_{\mu})+
\partial_{\mu}\phi^{\dagger}(x_{\mu})
\partial^{\mu}\psi^{\dagger}(x_{\mu})\right.\]
\begin{equation}
\label{eq3.13}
\left.-\frac {n+1} {2}m^2\left[
\phi^n(x_{\mu})\psi(x_{\mu})+
\phi^{n\dagger}(x_{\mu})\psi^{\dagger}(x_{\mu})\right]\right\}d^4x.
\end{equation}
Now we define i)  the free action ${\cal S}_0$
and ii) that corresponding to the interaction ${\cal S}_I$
as
\begin{equation}
\label{eq3.14}
{\cal S}_0=\int\left[
\partial_{\mu}\phi(x_{\mu})
\partial^{\mu}\psi(x_{\mu})+
\partial_{\mu}\phi^{\dagger}(x_{\mu})
\partial^{\mu}\psi^{\dagger}(x_{\mu})\right]d^4x,
\end{equation}
\begin{equation}
\label{eq3.15}
{\cal S}_I=-\frac {n+1} {2}m^2
\int\left[
\phi^n(x_{\mu})\psi(x_{\mu})+
\phi^{\dagger n}(x_{\mu})\psi^{\dagger}(x_{\mu})\right]d^4x.
\end{equation}
The fields in the interaction representation
satisfy the equations of motion for
free fields, corresponding to the action ${\cal S}_0$.
This is to satisfy the usual massless Klein-Gordon equation.
As a consequence, we can cast the fields
$\phi$ and $\psi$ in the fashion
\begin{equation}
\label{eq3.16}
\phi(x_{\mu})=\frac {1} {(2\pi)^{\frac {3} {2}}}
\int\left[\frac {a(\vec{k})} {\sqrt{2\omega}}
e^{ik_{\mu}x_{\mu}}+
\frac {b^{\dagger}(\vec{k})} {\sqrt{2\omega}}
e^{-ik_{\mu}x_{\mu}}\right]d^3k,
\end{equation}
\begin{equation}
\label{eq3.17}
\psi(x_{\mu})=\frac {1} {(2\pi)^{\frac {3} {2}}}
\int\left[\frac {c(\vec{k})} {\sqrt{2\omega}}
e^{ik_{\mu}x_{\mu}}+
\frac {d^{\dagger}(\vec{k})} {\sqrt{2\omega}}
e^{-ik_{\mu}x_{\mu}}\right]d^3k,
\end{equation}
where $k_0=\omega=|\vec{k}|$
The quantification of these two fields is i) immediately tractable  and
ii) the usual one, given by
\[[a(\vec{k}),a^{\dagger}(\vec{k^{'}})]=[b(\vec{k}),b^{\dagger}(\vec{k^{'}})]=
[c(\vec{k}),c^{\dagger}(\vec{k^{'}})]=\]
\begin{equation}
\label{eq3.18}
[d(\vec{k}),d^{\dagger}(\vec{k^{'}})]=
\delta(\vec{k}-\vec{k^{'}}).
\end{equation}
The naked propagator corresponding to
both fields is the customary one, and it is just  the Feynman propagator for massless fields
\begin{equation}
\label{eq3.19}
\Delta_0(k_{\mu})=\frac {i} {k^2+i0}
\end{equation}
where $k^2=k_0^2-\vec{k}^2$. 
The dressed propagator, which takes into account
the interaction, is given by
\begin{equation}
\label{eq3.20}
\Delta(k_{\mu})=\frac {i} {k^2+i0-i\Sigma(k_{\mu})},
\end{equation}
where $\Sigma(k_{\mu})$ is the self-energy.
Let us calculate the self-energy for the field
$\phi$ at second order in perturbation theory, for which the only non vanishing 
diagram corresponds to the 
 convolution of $n-1$ propagators for the field
$\phi$ and one  propagator for the field $\psi$. All remaining diagrams are null.
(this is easily demonstrated using the regularization
of Guelfand for integrals containing powers of
$x$ \cite{tp5}). Therefore, we have for the self-energy the
expression
\begin{equation}
\label{eq3.21}
\Sigma(k_{\mu})=\frac {(n+1)^2m^4} {4}\left(
\frac {i} {k^2+i0}\ast\frac {i} {k^2+i0}\ast\frac {i} {k^2+i0}
\cdot\cdot\cdot\ast\frac {i} {k^2+i0}\right).
\end{equation}
The convolution of $n$ Feynman's propagators 
of zero mass is calculated directly using the theory
of convolution of Ultradistributions \cite{tp6}-\cite{tp9}.
Here, we just give the result, that turns out to be  rather simple. A detailed demonstration  lies beyond this paper's scope. We arrive at 
\[\frac {i} {k^2+i0}\ast\frac {i} {k^2+i0}\ast\frac {i} {k^2+i0}
\cdot\cdot\cdot\ast\frac {i} {k^2+i0}=\]
\begin{equation}
\label{eq3.22}
\frac {i\pi^{2(n-1)}k^{2(n-2)}}
{\Gamma(n)\Gamma(n-1)}\left[
\ln(k^2+i0)+2\lambda(1)-\lambda(n-1)-\lambda(n)\right],
\end{equation}
where $\lambda(z)=\frac {d\ln\Gamma(z)}{dz}$.
The self-energy is then
\begin{equation}
\label{eq3.23}
\Sigma(k_{\mu})=
\frac {(n+1)^2m^4} {4}
\frac {i\pi^{2(n-1)}k^{2(n-2)}}
{\Gamma(n)\Gamma(n-1)}\left[
\ln(k^2+i0)+2\lambda(1)-\lambda(n-1)-\lambda(n)\right]
\end{equation}
For both fields $\phi$ and $\psi$, the self-energy
and the dressed propagator  coincide  up to second order.  

\vskip 3mm  \nd

Note that the current of probability is given by
\begin{equation}
\label{eq3.24}
{\cal J}_\mu=\frac {i} {4m}[\psi\partial_\mu\phi-\phi\partial_\mu\psi+
\phi^{\dagger}\partial_\mu\psi^{\dagger}-\psi^{\dagger}\partial_\mu\phi^{\dagger}].
\end{equation}
and it is  verified that
\begin{equation}
\label{eq3.25}
\partial_\mu{\cal J}^\mu=0
\end{equation}
This implies that the fields defined in the representation
of interaction are physical fields.

\setcounter{equation}{0}

\section{Conclusions}

We have here obtained some results that may be
regarded as interesting.

\nd 1) We developed the CFT for the particular q-SE
advanced in \cite{tp3}.

\nd 2) For this CFT we showed that the customary
dispersion relations apply. We also introduced the
physical fields i.e., those that the probability current
is conserved.
The physical states are introduced via analogy
with bosonic string theory.

\nd 3) We developed the QFT associated to the 
q-SE of \cite{tp3}. For weak fields, this q-QFT
coincides with the ordinary SE-QFT.
This result confirms our Nuclear Physics A results. These 
show that one needs energies of up to
1 TeV in order to clearly distinguish between
q-theories and q=1, ordinary ones.

\nd 4) Using Distribution Theory \cite{tp5} we
discussed the convolution of two Schr\"{o}dinger 
propagators obtaining a finite result.

\nd 5) We developed the QFT associated to the
q-KGE, generalizing our result of \cite{ts5}.

\nd 6) For low energies and q close to 1 this
theory coincides with the ordinary KG-QFT.

\nd 7) For particular q-values $q=\frac {n+1} {2}$
n integer we develop the q-KG-QFT.

\nd 8) We calculate the convolution of n naked
propagators, the corresponding self-energy
up to second order and the dressed propagator.
This was achieved  appealing to Ultradistributions
theory.

\end{document}